\input{aipcheck}

\documentclass[
    ,final            % use final for the camera ready runs
  ]
  {aipproc}

\usepackage{graphicx}

\usepackage{array}
\usepackage{amsmath}
\usepackage{amsfonts}
\usepackage{color}

\layoutstyle{6x9}

\begin{document}

\title{Realistic Simulation of Local Solar Supergranulation}



\author{Sergey D. Ustyugov}{
  address={4, Miusskaya sq., Keldysh Institute of Apllied Mathematics, Moscow, 125047, Russia}
}

\begin{abstract}

I represent results three-dimensional numerical simulation of solar surface 
convection on scales local supergranulation with realistic model physics. 
I study thermal structure of convective motions in photosphere, the range of 
convection cell sizes and the penetration depths of convection. A portion of the 
solar photosphere extending 100 x 100 Mm horizontally and from 0 Mm down to 20 Mm 
below the visible surface is considered. I take equation of state and opacities of 
stellar matter and distribution with radius of all physical variables from Solar 
Standard Model. The equations of fully compressible radiation hydrodynamics with 
dynamical viscosity and gravity are solved. The high order conservative PPML 
difference scheme for the hydrodynamics, the method of characteristic for the 
radiative transfer and dynamical viscosity from subgrid scale modeling are applied. 
The simulations are conducted on a uniform horizontal grid of 1000 x 1000, with 168 
nonuniformly spaced vertical grid points, on 256 processors with distributed memory 
multiprocessors on supercomputer MVS5000 in Computational Centre of Russian Academy of 
Sciences.

\end{abstract}

\maketitle

%%%%%%%%%%%%%%%%%%%%%%%%%%%%%%%%%%%%%%%%%%%%
%% MAINMATTER
%%%%%%%%%%%%%%%%%%%%%%%%%%%%%%%%%%%%%%%%%%%%

%\section{Introduction}

We conducted numerical simulation development of solar convection on scales of 
mesogranulation and supergranulation in a three-dimensional computational box. 
In previous work by the author \cite{Ustyugov:2006} it was shown that there is
collective motion of small-scale and short-lived granules from centre to edges of 
supergranular cells. The similar idea was suggested also in article \cite{Rast:2004}. 
Simulation of solar photosphere convection \cite{Stein:2006} in a computational 
domain of size 48 Mm in horizontal plane and on depth 20 Mm showed that the sizes 
of convective cells increase with depth. The purpose of this work is to investigate 
the development and scales of convection in a region of size 100 Mm in the horizontal 
plane and on depth 20 Mm. 

%\section{Numerical method}

We take distribution of the main thermodynamic variables  by radius
due to Standard Solar Model \cite{Christensen-Dalsgaard:2003} with parameters
$(X,Z,\alpha)=(0.7385,0.0181,2.02)$, where $X$ and $Y$ are hydrogen
and helium abundance by mass, and $\alpha$ is the ratio of mixing
length to pressure scale height in convection region. We use OPAL tables
for opacities and equations of state for solar matter from work\cite{IglesiasRogers:1996}.
For solution of equations of compressible hydrodynamics 
we apply our new method: piecewise parabolic method on local stencil (PPML) 
\cite{Popov:2007}. In order to evaluate effect of subgrid scales on evolution 
of convective motion we use Large Eddy Simulations technique. Rate dissipation 
was defined taking into account action buoyancy force and shear motions  
\cite{Canuto:1994}. The one step of time integration for source terms is defined 
by third order TVD Runge-Kutta method \cite{ShuOsher:1988}. The scheme is second order 
by space and time. For approximation viscous terms we used central differences. 
For evaluation radiative term in energy equation we used method of short characteristic 
for upper part of domain and diffusion approximation for bigger depths. 
We use uniform grid in x-y directions and nonuniform grid in vertical z one. 
Boundary conditions in horizontal planes is periodic and in vertical plane is 
characteristic. An initial moment we deposit to domain small disturbance of velocity with 
maximum of amplitude $10^{-4}$.

%\section{Results}

The numerical simulation of local solar supergranulation was conducted  
during 48 solar hours. Near the solar surface in the horizontal plane
(Figure 1a) we find signs of small-scale turbulent convection. 
Average sizes of cells is 1-1.5 Mm and lifetime of ones is 1-2 min. 
(Figure 1a-1b).  With increasing of depth the size of convective cells grows. 
At depth 2-5 Mm the average size of a cells is 10-15 Mm (Figure 1c-1e), 
whereas at depths bigger than 10 Mm it is 20-30 Mm (Figure 1f). At the vertices 
of convective cells at depths 3-4 Mm the motion of material becomes supersonic. 
In the deepest layers of photosphere the flow of material becomes slower and more laminar. 
In a vertical plane we find three different regions of evolution convection 
(Figure 2,3). At depths 0-2 Mm, the small-scale turbulent motion of material is formed.
In this region the action of strong compressible effects on sides of granules 
leads to movement of cold material to down. In some points there is strong vorticity 
motion where material moves from different directions to one point. There we find 
formation of more powerful downdrafts, that moves to down with maximum of velocity.  
As in previous investigation \cite{Ustyugov:2008} we find transition zone at depths 
of 2-5 Mm. In this region convection develops on scales mesogranulation. At depths 
of 4-5 Mm the downdrafts has a maximum velocity, in average, of about 4 km/sec 
with value of Mach number 1.5. Material moves downwards in the form of narrow jets 
and upwards in wider regions on size of mesogranulation. In the third part of the 
computational domain, on depths bigger 15 Mm, we discover places where material 
continues movement to down and reaches the bottom boundary with subsonic velocity. 
The distance between these flows is 20-30 Mm and defines development of the 
convective cells on the scale of supergranulation.

%\section{Acknowledgments}

I would like to thank Dr. Christiana Dumitrache from Bucharest Observatory 
and Local Organize Committee for financial support for my participation 
in conference "Exploring the Solar System and the Universe".

\begin{figure}[h+]
  \includegraphics[angle=-90,scale=0.32]{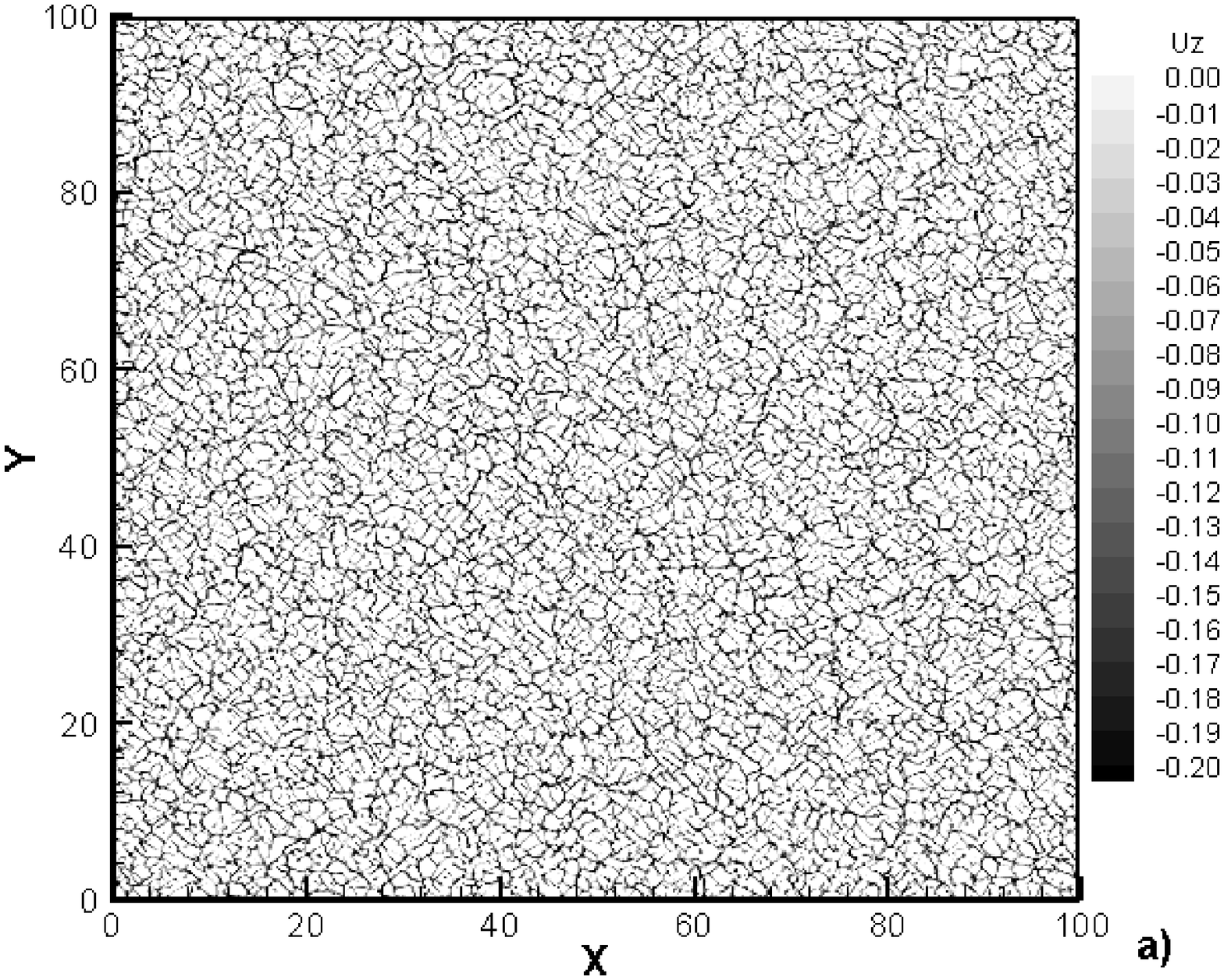}
  \includegraphics[angle=-90,scale=0.32]{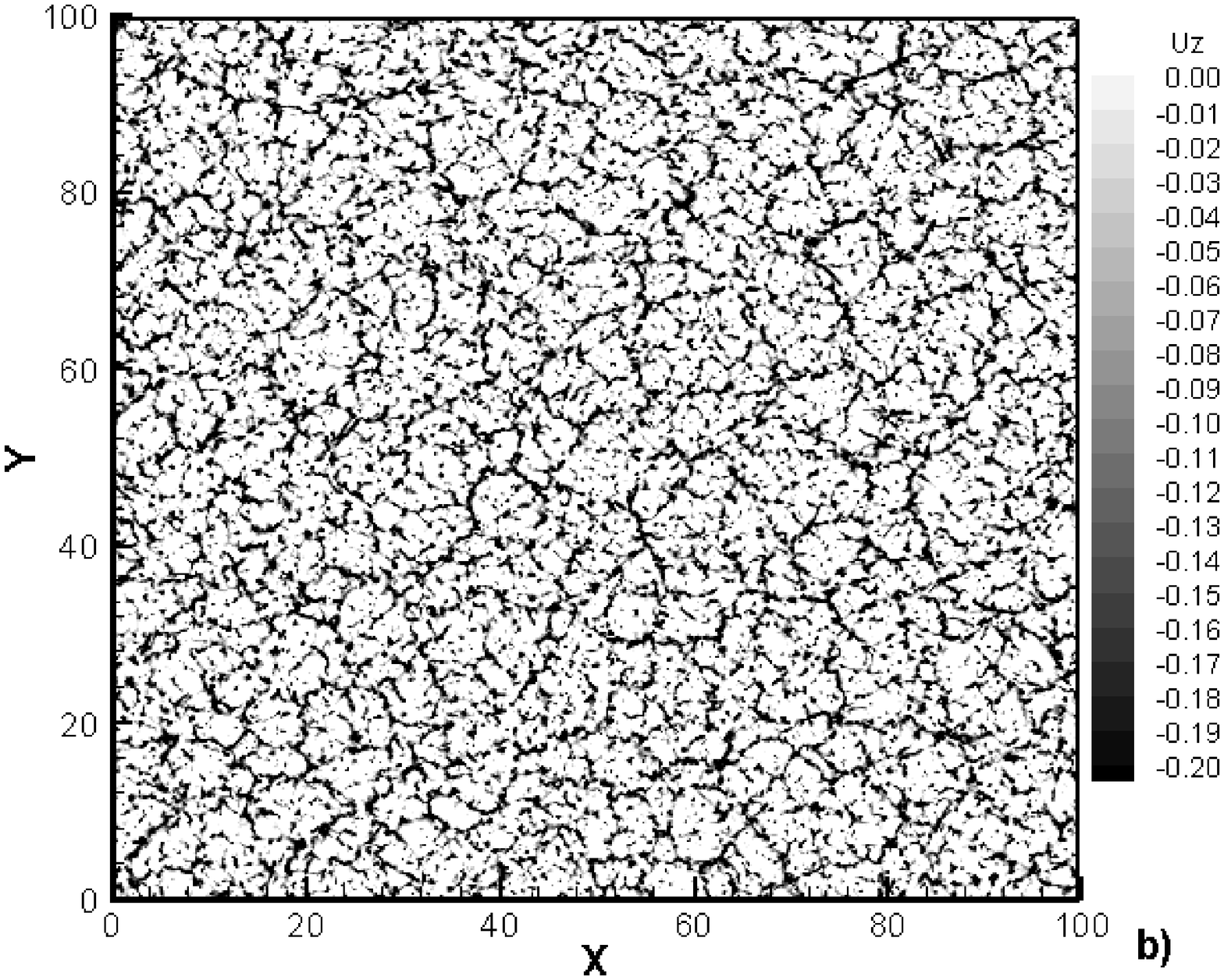}
\end{figure}

\begin{figure}[h+]
  \includegraphics[angle=-90,scale=0.32]{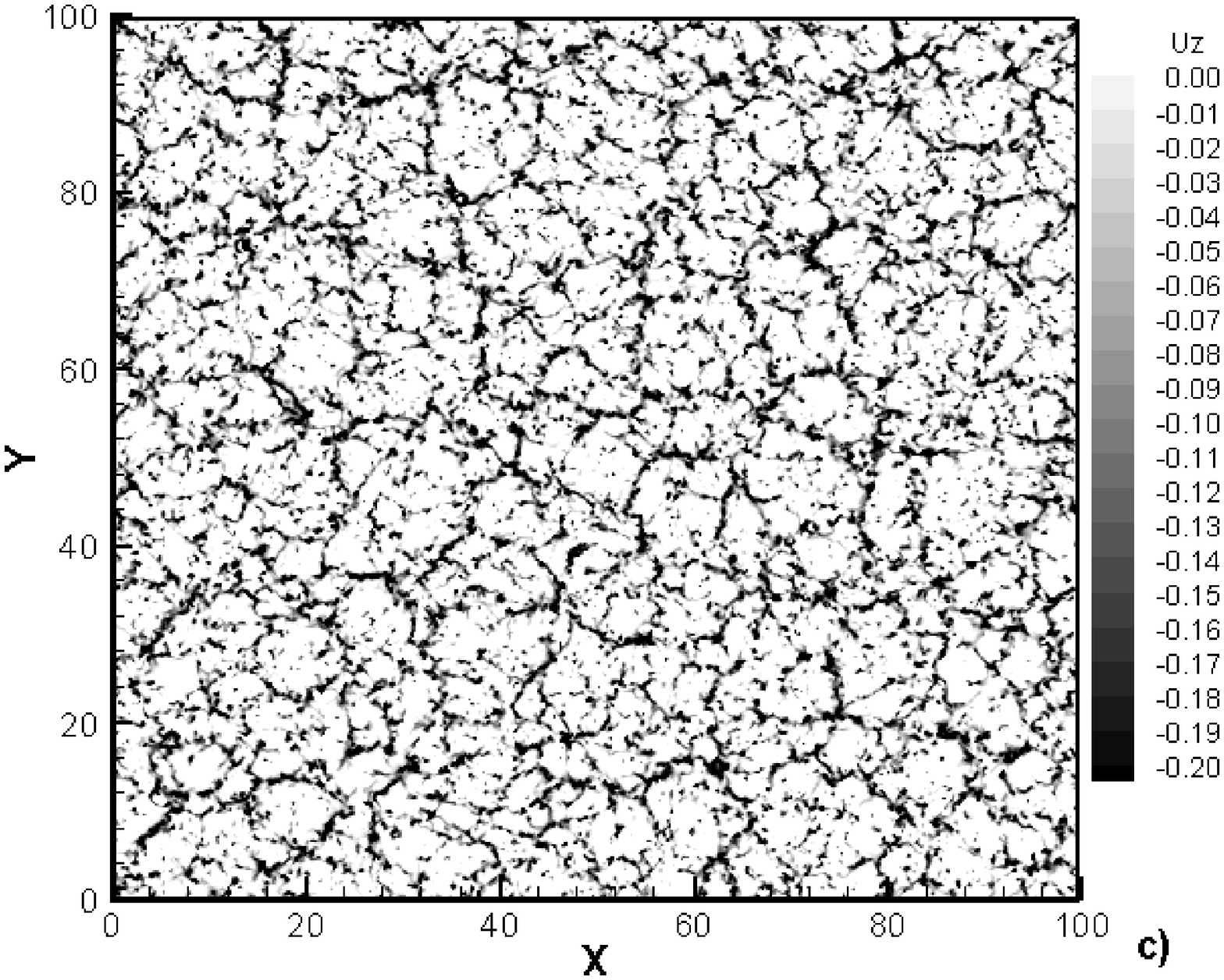}
  \includegraphics[angle=-90,scale=0.32]{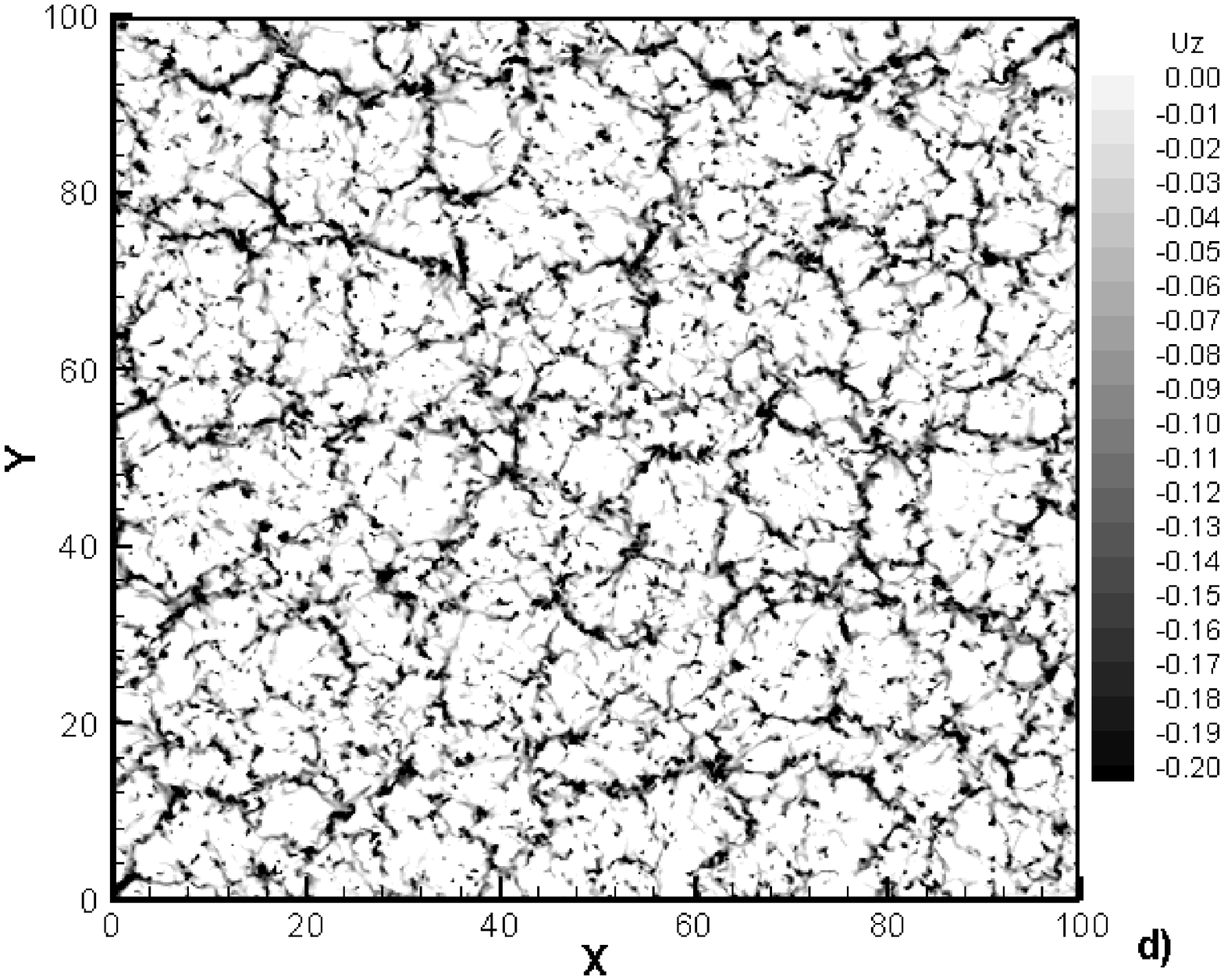}
\end{figure}

\begin{figure}[h+]
  \includegraphics[angle=-90,scale=0.32]{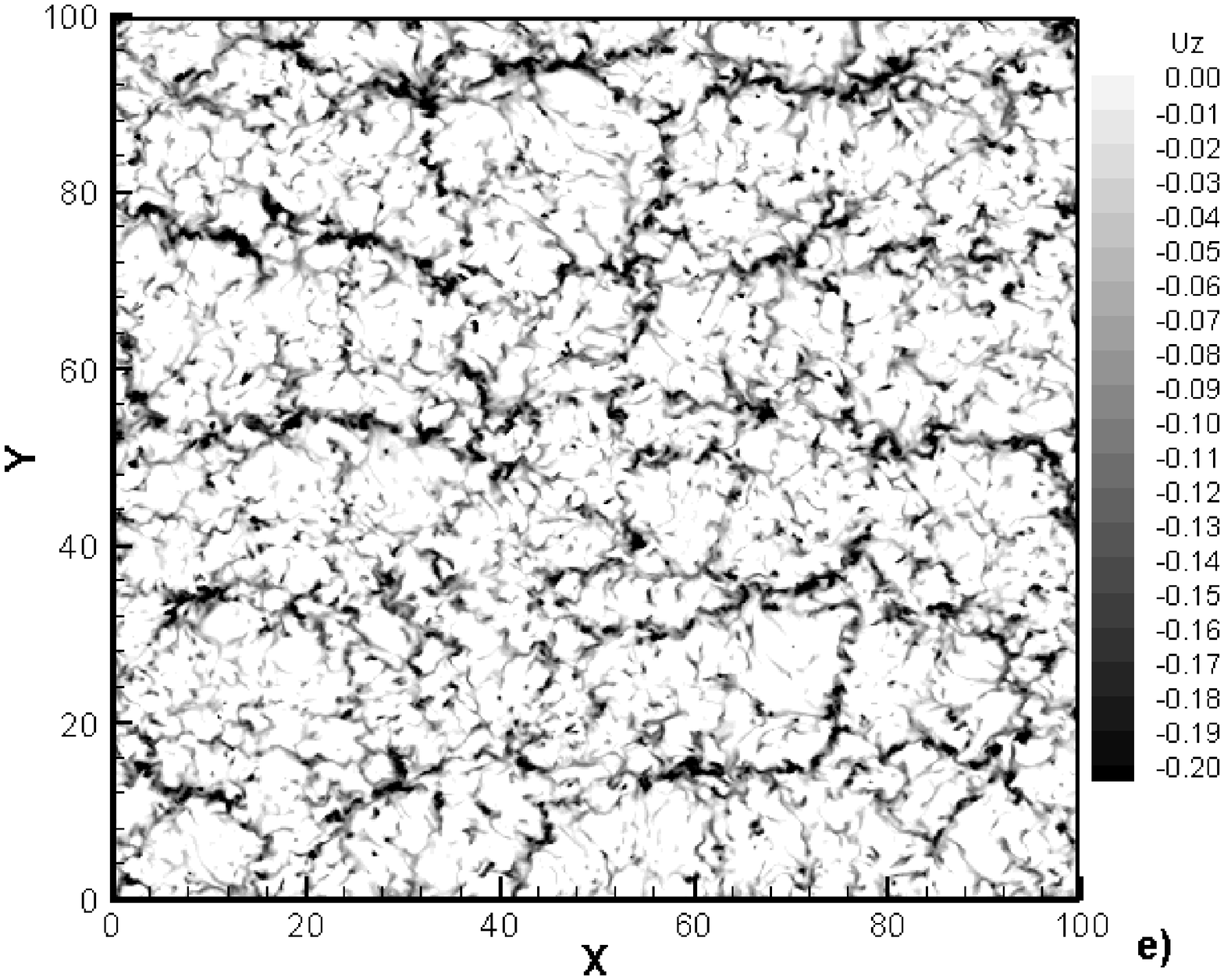}
  \includegraphics[angle=-90,scale=0.32]{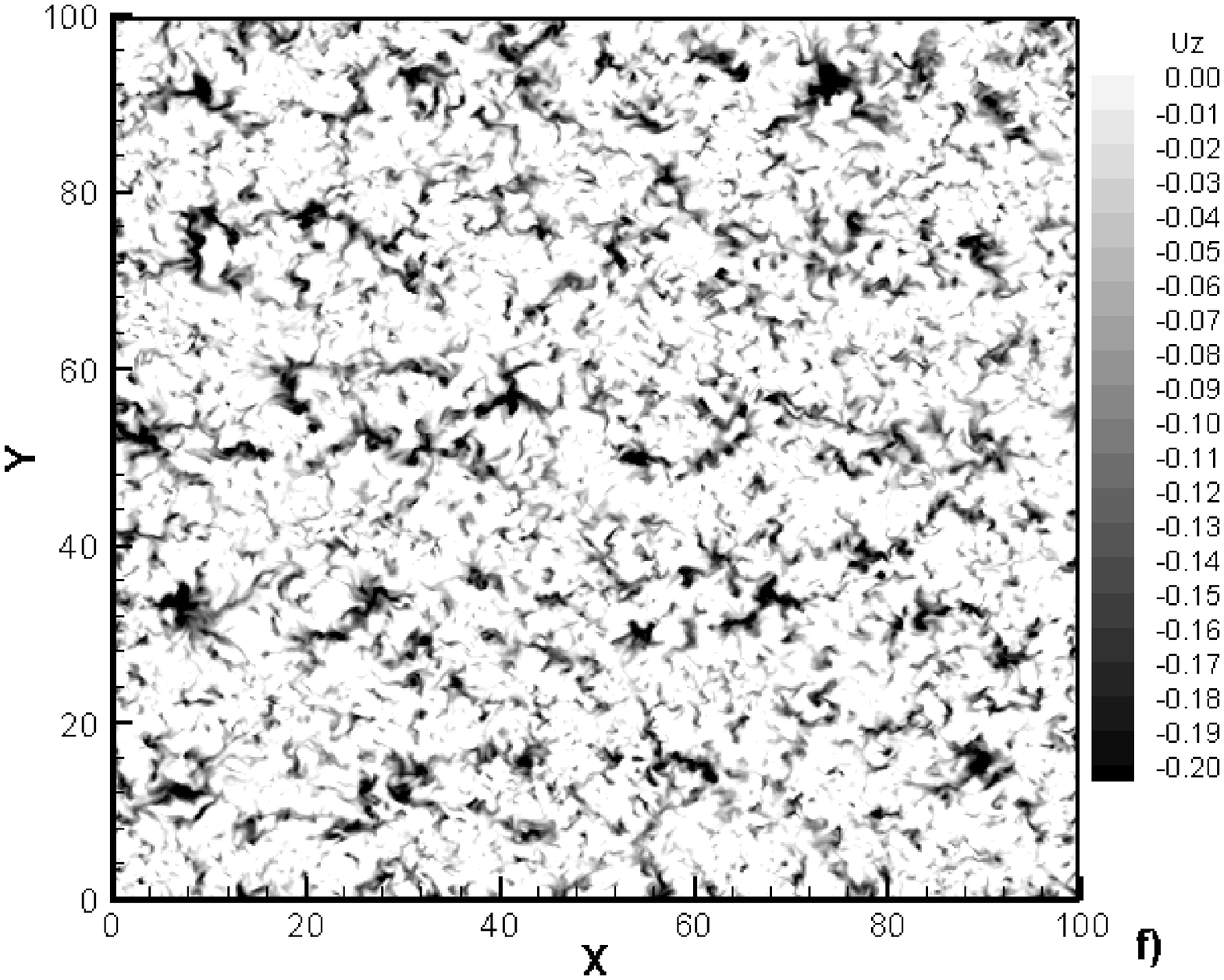}
\caption{Contours of the vertical component of velocity (in units speed of sound) 
at depths 0 ({\it a}), 1 ({\it b}), 2 ({\it c}), 3 ({\it d}) 5 ({\it e}), 10 ({\it f}) Mm.
The X and Y axises are scaled in Mm.}
  \end{figure}

\begin{figure}[h+]
 \includegraphics[angle=-90,scale=0.65]{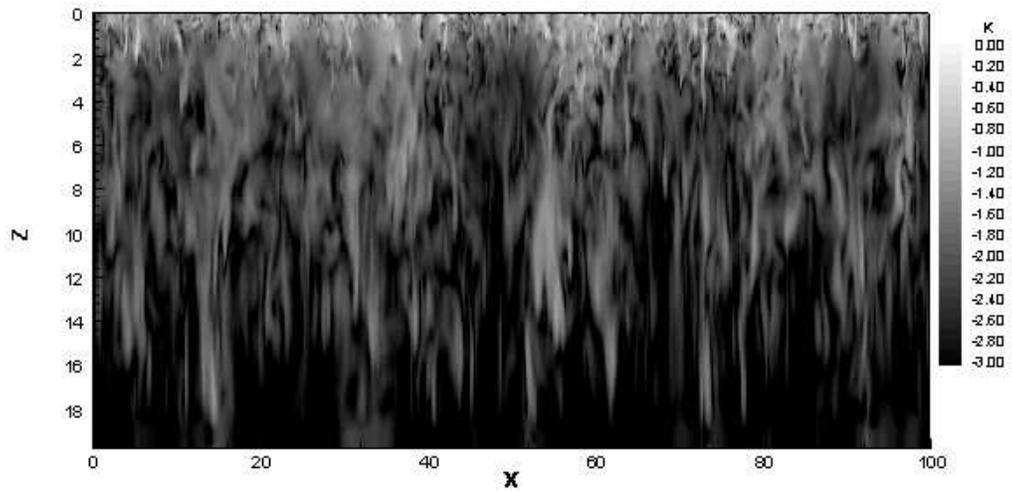}
  \caption{Image of contours of the logarithm of kinetic energy in vertical plane.
  The X and Y axises are scaled in Mm.}
\end{figure}

\begin{figure}[h+]
 \includegraphics[angle=-90,scale=0.65]{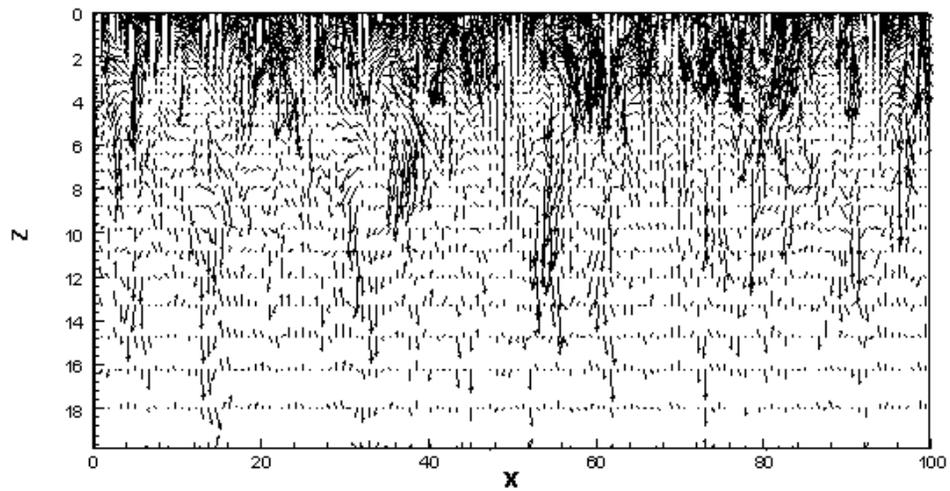}
  \caption{Image of contours of the field of velocity in vertical plane.
  The X and Y axises are scaled in Mm.}
\end{figure}
\end{document}